\documentclass[prb,aps,twocolumn,floats,showpacs]{revtex4}
\usepackage{amssymb}

\usepackage{epsfig}

\newcommand{\be}{\begin{equation}}
\newcommand{\ee}{\end{equation}}
\newcommand{\bea}{\begin{eqnarray}}
\newcommand{\eea}{\end{eqnarray}}

\begin{document}
\title{Conductance of a single-atom carbon chain with graphene leads}
\author{Wei Chen}
\affiliation{ Department of Physics, University of Washington,
Seattle, WA 98195-1560}
\author{A. V. Andreev} \affiliation{
Department of Physics, University of Washington, Seattle, WA
98195-1560}

\author{G.~F. Bertsch}
\affiliation{Institute of Nuclear Theory, Department of Physics,
University of Washington, Seattle, WA 98195-1560}
\date{\today}

\begin{abstract}
We study the conductance of an interconnect between two graphene
leads formed by a single-atom carbon chain. Its dependence on the
chemical potential and the number of atoms in the chain is
qualitatively different from that in the case of normal metal leads.
Electron transport proceeds via narrow resonant states in the wire.
The latter arise due to strong reflection at the junctions between
the chain and the leads, which is caused by the small density of
states in the leads at low energy. The energy dependence of the
transmission coefficient near resonance is asymmetric and acquires a
universal form at small energies. We find that in the case of leads
with the zigzag edges the dispersion of the edge states has a
significant effect on the device conductance.
\end{abstract}

\maketitle

\section{Introduction }
\label{sec:intro}

The efforts to miniaturize electronic devices have long motivated
studies of electron transport in molecular and atomic scale devices.
Creating reliable electrical contacts with the molecule presents a
major challenge in molecular electronics. Until recently, in
molecular electronic devices the molecule was typically attached
between two normal metal, such as gold,
electrodes~\cite{Reed1997,Reed1999,Park2000,Park2002,Liang2002,AtomicContacts}.

Carbon-based conductors have long been expected to be promising
components of electronic devices~\cite{McEuen1998}. Apart from bulk
graphite there  are also quasi-one dimensional (carbon
nanotubes~\cite{Iijima}) and two-dimensional
(graphene~\cite{graphene_Geim}) forms, which have remarkable
mechanical and electrical properties and form strong chemical bonds
with each other. This offers the  prospect of building entire
electronic devices or circuits out of carbon-based materials.

Single-atom carbon chains (SACCs) are natural components of such
devices. They are expected to be ideal one-dimensional
conductors~\cite{Avouris}. They covalently bond to other carbon
materials. Formation of SACCs was conjectured to occur between
carbon nanotubes~\cite{Richter,NanoLettersChains,Louie} and in gaps
between two graphene leads~\cite{Marc}. Formation of SACCs between
graphene electrodes fabricated by stretching a graphene strip has
been observed in Ref.~\onlinecite{Jin2009}. SACC interconnects
between graphene leads could form a basic unit for integration into
more complicated circuits in the future. Electron transport through
SACCs with graphene contacts has not been studied theoretically. We
address this issue in the present paper.

Electron transport in SACCs with metal leads has been studied
numerically~\cite{Avouris,Larade2001,Tongay2004,Crljen2007}. The
conductance of an SACC strongly coupled to metal electrodes is of
order of the conductance quantum and exhibits even-odd oscillations
with the number of atoms in the chain, with the contrast ratio of
order unity.

In this paper we show that electron transport through an SACC
interconnect between graphene leads in the non-interacting electron
approximation is described by an analytically solvable model. This
enables us to gain physical insight into the essential features of
electron transport. This model can also serve as the starting point
for treating one-dimensional electron correlations in SACC and their
influence on transport.

The conductance of the system is qualitatively different from that
in the case of metal leads. For all electron energies corresponding
to practically relevant temperatures and doping levels the junction
between the chain and the graphene lead is almost perfectly
reflecting even at strong coupling between the chain and the lead.
The transmission coefficient of the contact vanishes linearly with
electron energy as the latter approaches the Fermi energy of undoped
graphene. This suppression of transmission results from the
vanishing of density of states (DoS) in graphene at zero doping. As
a result electron transport through the interconnect proceeds via
narrow resonant states in the chain that arise due to strong
reflection at the junctions. The width and the position of the
resonances depend on the length of the interconnect and the details
of its coupling to the leads.  The shape of low energy resonances is
universal but markedly different from the Breit-Wigner form. It is
dictated by the linear energy dependence of the DoS in graphene at
the point of contact with the chain. This holds even in the case of
leads with zigzag edges, which support edge
states~\cite{edge_1996,prb1996}. Although edge states have a linear
dispersion at low energies, their wave functions extend into the
bulk to distances inversely proportional to the energy and give a
linear in energy contribution to DoS at the contact point.

Due to the resonant character of transmission the device conductance
is very sensitive to the number of atoms in the chain. In the case
of graphene leads the conductance difference between chains with odd
and even number of atoms in the SACC (which was first
noted~\cite{Avouris} in the case of metal leads) becomes much more
pronounced and appears as the difference between the on- and off-
state conductance.

The high stability of SACC interconnects makes them promising
building block of atomic scale electronics of the future. The
resonant character of electron transport through them suggests that
they can be used as components of atomic scale transistors. Our work
is the first step towards theoretical understanding of electron
transport in SACC interconnects between graphene leads.

The paper is organized as follows. In Sec.~\ref{sec:qualitative} we
qualitatively discuss the essential features of electron transport
in SACC interconnects between graphene leads. In
Sec.~\ref{sec:system and model} we formulate an analytically
solvable model of electron dynamics in the system. In
Sec.~\ref{sec:contact} we derive a general formula for the
reflection amplitude of the junction between the chain and the lead
in terms of the electron Green function (GF) of the lead. In
sec.~\ref{sec:lead} we evaluate the GF and study the tunneling
density of states at the junction due to the bulk and edge electron
states in the lead. In sec.~\ref{sec:conductance_wire} we evaluate
the transmission coefficient of the device and obtain the universal
formula for the resonance shape. We discuss our results and
experimental implications in Sec.~\ref{sec:discussion}.

\section{Qualitative discussion }
\label{sec:qualitative}

Linear molecules with degenerate electron states represent an
exception to the Jahn-Teller theorem on the instability of the
symmetric molecular configurations~\cite{Landau_QM}. In this case
the degenerate electron states have nonzero angular momenta $\pm
\Lambda$ about the molecule axis. Thus the matrix element coupling
the two states and the corresponding energy gain arises only in
second or higher order in the vector displacements of the nuclei
from the symmetry axis. In the linear SACC the two degenerate
electron bands have angular momenta $\pm 1$ about the molecular
axis, and their coupling arises only in second order in the nuclear
displacements. In the Peierls channel, which at zero doping
corresponds to dimerization,  the coupling between right and left
movers is linear in the displacements. Thus the two most likely
candidates for the SACC structure are the linear chain of
equidistant double bonded carbon atoms, known as cumulene ($\cdots
\mathrm{C}=\mathrm{C}=\mathrm{C}=\mathrm{C} \cdots$), and the
Peierls-distorted dimerized chain, known as polyyne ($\cdots
\mathrm{C}-\mathrm{C} \equiv \mathrm{C}-\mathrm{C} \cdots$). This
expectation is confirmed by numerical studies. \emph{Ab initio}
calculations~\cite{Tongay2004} show that because of large quantum
fluctuations of the atomic positions, among all possible spatial
arrangements of the carbon atoms in the chain only the cumulene
structure is stable. The outer shell electron orbitals in cumulene
are $sp$-hybridized. The two $sp$ hybridized orbitals form the fully
occupied $\sigma$ band. The two remaining $p_y$ and $p_z$ orbitals
form two doubly degenerate $\pi$ bands, which are half filled making
undoped cumulene a one-dimensional conductor. Numerical
studies~\cite{Okano2007} also show that cumulene remains metallic
under doping.

If the SACC attached to the graphene leads is not too long the main features of
electron transport through it can be understood within the single electron
picture without accounting for electron-electron correlations.

As the conduction $\pi$ band in graphene leads is formed by the
$p_z$ orbitals, which are perpendicular to the graphene plane, only
the electrons from the $\pi_z$ band in SACC can propagate into the
leads. Thus electron transport is mediated by a single conducting
band in the chain.

Because of the long mean free path of electrons in
graphene~\cite{graphene_Geim}, electron motion in the leads may be
assumed to be ballistic. The device conductance is then determined
by the elastic electron scattering at the junctions between the
molecular wire and the leads and backscattering in the chain.
Backscattering in the SACC due to electron-phonon interaction
corresponds to emission or absorption of phonons with a wavelength
of order of the interatomic spacing and energy of the order of $10^3
K$. As a result such processes are exponentially
suppressed~\cite{Seelig1,Seelig2} even at room temperature. Although
imperfections in the substrate and deviations of the atomic
positions from the ideal configuration cause some backscattering of
electrons in the wire, we show below that the strongest reflection
of the electron wave in the chain occurs at the contact with the
graphene lead.

With the aid of the Fermi golden rule the transmission coefficient
of the contact can be estimated as $\mathcal{T}_c \sim | \gamma_c|^2
\nu_w \nu_g$. Here $\gamma_c$ is the tunneling matrix element that
couples the last atom in the chain to the lead, $\nu_w$ is the local
DoS at that atom and $\nu_g$ the DoS an the graphene atom that is
connected to the chain. The local DoS in the chain is energy
independent and can be estimated as $\nu_w\sim 1/\gamma_w$, where
$\gamma_w$ is the nearest neighbor hopping integral in the atomic
wire. The DoS in graphene, on the other hand is strongly energy
dependent and is of the order of $\nu_g \sim |\epsilon|
/\gamma_g^2$, where $\gamma_g$ is the nearest neighbor hopping
integral in graphene and $\epsilon$ is the electron energy measured
from the Fermi level of undoped graphene~\footnote{It might seem
that the presence of edge states might give rise to an
energy-independent contribution to the tunneling DoS at the point.
This is not so however. The wave functions of edge states at low
energies extend into the bulk to distances inversely proportional to
the electron energy. As a result their contribution to the local DoS
at the contact is also linear in $\epsilon$. It is analyzed in
detail in section \ref{sec:tunneling_DOS}}. Thus the transmission
coefficient of the contact can be estimated as
\begin{equation}\label{eq:contact_transmission_estimate}
    \mathcal{T}_c \sim \frac{\gamma^2_c|\epsilon|}{\gamma_w\gamma_g^2}.
\end{equation}
The hopping integrals in graphene and SACC are of the same order of magnitude.
Therefore at typical doping levels, $|\epsilon| \ll \gamma_g$, the transmission
coefficient is small even at strong coupling between the chain and the lead, when
all hopping integrals between nearest neighbor carbon atoms are of the same
order, $\gamma_c \sim \gamma_w\sim \gamma_g $.

Neglecting the weak backscattering in the wire we can express the
energy-dependent transmission coefficient of the device, $\mathcal{T}(\epsilon)$,
in terms of the reflection amplitudes of the left and right junction between the
wire and the leads, $r_{l/r}(\epsilon)=|r_{l/r}(\epsilon)|\exp (i \delta_{l/r})$,
as
\begin{equation}\label{eq:transmission_r}
    \mathcal{T}(\epsilon)={|t_l(\epsilon)|^2 |t_r(\epsilon)|^2\over
(1-|r_l(\epsilon)||r_r(\epsilon)|)^2+2 |r_l(\epsilon)||r_r(\epsilon)| (1-\cos
\phi)}.
\end{equation}
Here $|t_{l/r}(\epsilon)|^2=1-|r_{l/r}(\epsilon)|^2$ are the transmission
coefficients of the junctions, and $\phi$ is the phase accumulated by an electron
upon returning to the same point in the chain after being reflected from both
contacts. It can be expressed as $\phi=2k \mathcal{N}+\delta_l +\delta_r$, where
$\mathcal{N}$ is the number of atoms in the chain, $k$ is the absolute value of
the dimensionless (measured in units of the inverse lattice spacing $d$ of the
chain) electron quasimomentum, and $\delta_{l/r}$ are the phases of the
reflection amplitudes of the contacts.

Because at low energies the junctions become strongly reflective appreciable
transmission through the device in this regime occurs only near resonances, where
the phase $\phi$ equals an integer multiple of $2\pi$. The energy spacing between
adjacent resonances is $\sim \gamma_c/ \mathcal{N}$.

To obtain a simplified expression for the transmission coefficient near a low
energy resonance we write the reflection amplitudes of the junctions at low
energies as $|r_{l/r}(\epsilon)|=1-c_{l/r}|\epsilon|/\gamma_g$, where $c_{l/r}$
is a numerical coefficient of the order of unity. This expression follows from
Eq.~(\ref{eq:contact_transmission_estimate}). Linearizing the energy dependence
of the phase $\phi$ near the resonance energy $\epsilon_0$, $\phi= c_\phi (
\epsilon -\epsilon_0) /\gamma_g$, where $c_\phi \sim \mathcal{N}$ is a numerical
coefficient, we can write the transmission coefficient of the device as,
\begin{equation}\label{eq:transmission_resonance_asymmetric}
    \mathcal{T}(\epsilon)=\left[\mathcal{T}_0^{-1} + \alpha
    \left(1 -\frac{\epsilon_0}{\epsilon} \right)^2  \right]^{-1}.
\end{equation}
Here $\mathcal{T}_0=(c_l+c_r)^2/(4c_lc_r)$ is the transmission coefficient at the
resonance, and $\alpha =c_\phi^2/( 4 c_l c_r) \sim \mathcal{N}^2$.

The width of the resonance is of the order of
$\epsilon_0/\mathcal{N}$. In the case of symmetric contacts,
$c_l=c_r$, the device becomes perfectly transmitting on resonance.
The shape of the resonance is shown in  Fig.~\ref{fig:2}. It is
strongly asymmetric and markedly different from the Breit-Wigner
form which arises in the case of metal leads.

Since low energy transmission through the device proceeds via a
single resonant state in the chain it is clear that
Eq.~(\ref{eq:transmission_resonance_asymmetric}) holds under very
general conditions.  The assumption that the coupling $\gamma_c$
between the chain and the lead is energy independent is valid if the
resonance energy is smaller than the inverse propagation time of an
electron across the junction. Such resonances always exist if the
SACC  is longer than the junction (for example a small peninsular
extending between the graphene lead and the chain). As long as the
backscattering in the chain does not lead to localization the
resonant state will remain coupled to both leads. The backscattering
will merely modify its energy and strength of coupling to the leads,
and can be accounted for by the change of parameters in
Eq.~(\ref{eq:transmission_resonance_asymmetric}). Similarly
electron-electron interactions in a finite chain will renormalize
the energy and the coupling of the resonant state with the leads.

In the remainder of the paper we present a quantitative treatment of
simple model of electron transport through an cumulene SACC
interconnect between graphene leads.

\begin{figure}[ptb]
\includegraphics[width=7.0cm]{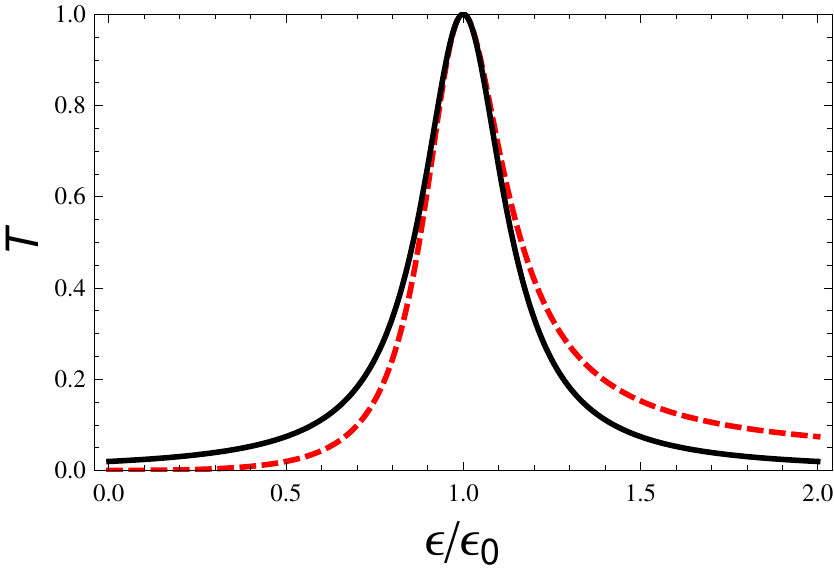}
\caption{Color online. Dependence of the transmission coefficient
$\mathcal{T}$ in Eq.~(\ref{eq:transmission_resonance}) on
$\epsilon/\epsilon_0$ (dashed line) with $N=10$ and $\alpha\sim 1$.
The solid curve is the Breit-Wigner resonance with the same
resonance energy $\epsilon_0$ and width. }\label{fig:3}
\end{figure}

\section{System and model}
\label{sec:system and model}

\begin{figure}[ptb]
\includegraphics[width=8cm]{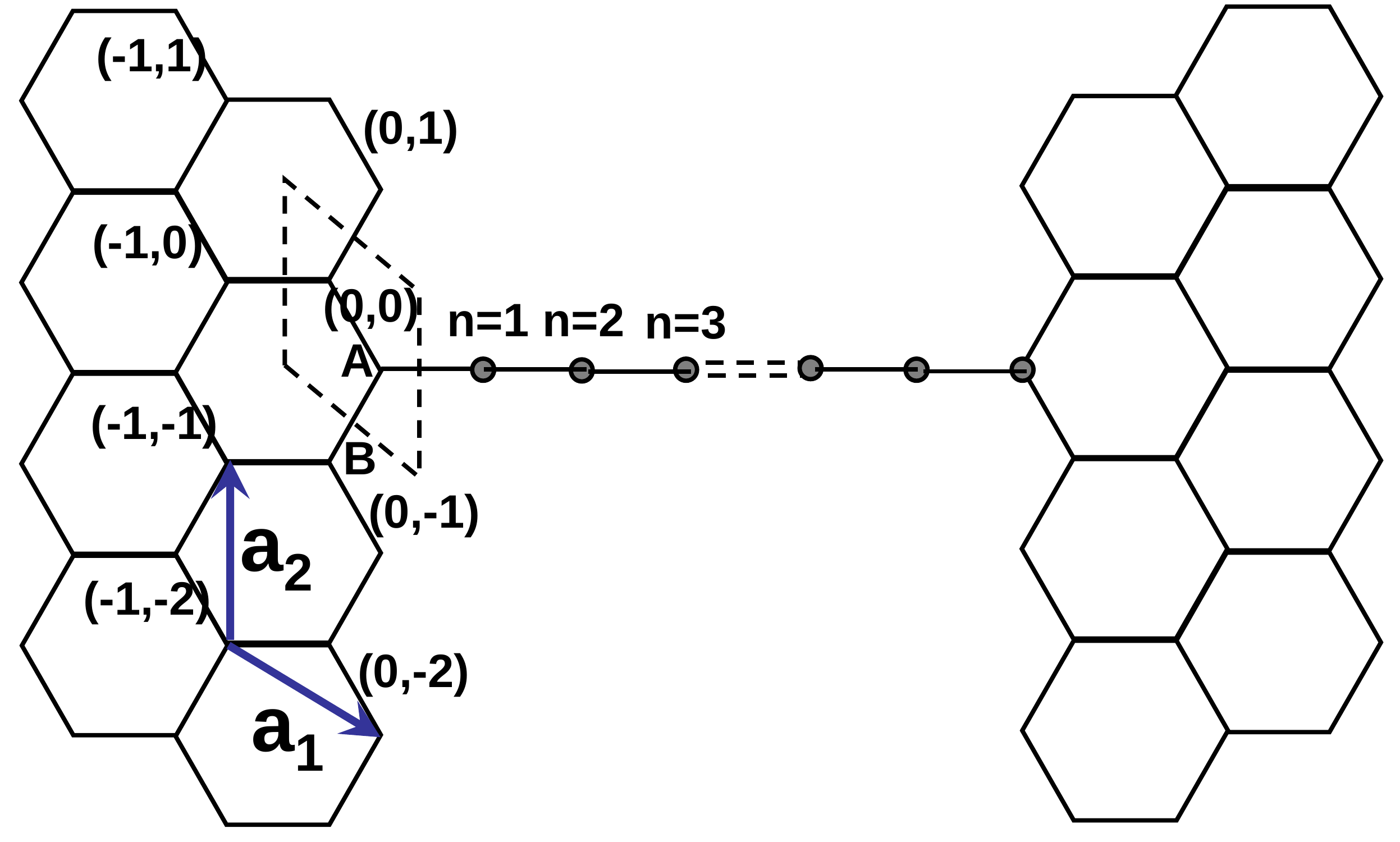}
\caption{Schematic picture of the device. The atoms in the wire are labeled by
$n=1,2,\ldots$. The unit vectors of the graphene Bravais lattice, $\mathbf{a}_1$
and $\mathbf{a}_2$, are shown by blue arrows. Each unit cell (dashed rhomboid) is
labeled by $(N_1, N_2)$.} \label{fig:1}
\end{figure}

Consider an ideal cumulene SACC connected to graphene leads with perfect zigzag
edges, as shown in Fig.~\ref{fig:1}

As a first step in the theoretical analysis of the system, we assume
that the atoms in the SACC are in the ideal cumulene configuration.
We work in the noninteracting electron approximation and describe
the electron motion in the conducting $\pi_y$ and $\pi_z$ bands
using the nearest neighbor tight binding approximation. More
complicated band structure of the carbon wire will not significantly
modify our conclusions.

Electron transport through the device is fully determined by the
reflection amplitude of the contact between the SACC and the
graphene lead. We derive in Sec.~\ref{sec:contact} a general formula
for the reflection amplitude of the contact in terms of the local
DoS in the lead evaluated at the atom which is connected to the
SACC, Eq.~(\ref{eq:reflection_G}). This expression is applies for an
arbitrary shape of the lead. Then in Sec.~\ref{sec:lead} we
specialize to the case of graphene leads with zigzag edge. The
zigzag edge is likely to be formed as a result of an electric
failure~\cite{Marc} or a tear of a graphene strip because it has the
least number of dangling bonds per unit length. We find that the
edge states present in the case of zigzag edge provide a significant
contribution to the tunneling DoS at the edge of the lead, and thus
carry a significant portion of the current through the device.

\subsection{Reflection amplitude of the junction}
\label{sec:contact}

Let us first consider a single junction between the SACC and a graphene lead. We
label the sites in the chain by an integer $n$ which enumerates the atoms
starting from the junction, see Fig.~\ref{fig:1}. The reflection amplitude of the
contact can be found from the retarded Green function of the auxiliary system
evaluated between two points inside the semi-infinite wire using the expression,
\begin{equation}\label{eq:r_G_wire}
    G(n,n')\sim \exp[-ik|n-n'|] + r(\epsilon)\exp[-ik(n+n')].
\end{equation}
Here  $k$ is the absolute value of the energy-dependent
quasimomentum of the electron in the chain, and the Green function
is defined in terms of the system Hamiltonian $\hat{H}$ in the
standard way,
\begin{equation}\label{eq:GF_def}
    G(n,n')=\langle n| \hat G|n'\rangle=
    \langle n| (\epsilon_+ -\hat{H})^{-1}|n'\rangle,
\end{equation}
where $\epsilon_+=\epsilon +i 0$.

We write Hamiltonian of the system as
\begin{equation}\label{eq:H_sum}
    \hat{H}=\hat{H}_w+\hat{H}_g+ \hat{V},
\end{equation}
where $\hat{H}_w$ and $\hat{H}_g$ are respectively the Hamiltonians of the
semi-infinite wire and the semi-infinite graphene lead, and $\hat{V}$ is the
perturbation, which describes electron tunneling between them.

Introducing the Green function of the unperturbed system , $\hat G_0
=(\epsilon_+ -\hat H_w - \hat H_g)^{-1}$, we can express the Green
function of the full system as
\begin{equation}\label{eq:G_T}
    \hat G=\hat G_0 +\hat G_0 \hat T \hat G_0,
\end{equation}
where $\hat T$ is the $T$-matrix of the junction between the chain
and the lead given by
\begin{equation}\label{eq:T}
    \hat{T}= (1-\hat{V}
\hat{G}_{0})^{-1}\hat{V}.
\end{equation}

In the nearest neighbor tight binding model the tunneling
perturbation $\hat V$ couples only the $| n=1 \rangle$ orbital in
the chain and a single contact site in the graphene lead, which we
label as $| \mathbf{0} \rangle$. In the $2 \times 2$ subspace
spanned by these states the tunneling perturbation $\hat{V}$ can be
written as
\begin{equation}\label{eq:V}
\hat V=\gamma_c\left(\begin{array}{cc}
  0 & 1 \\
  1 & 0
\end{array}\right),
\end{equation}
where $\gamma_c$ is the hopping integral at the contact between the
chain and the graphene lead. In this case the $T$-matrix depends
only on the unperturbed Green function within the $2\times 2$
subspace, where it has the form
\begin{equation}\label{eq:G_0}
\hat{G}_0=\left(\begin{array}{cc}
  G_g(\mathbf{0},\mathbf{0}) & 0 \\
  0 & G_w(1,1) \\
\end{array}\right).
\end{equation}
Here $G_g(\mathbf{0},\mathbf{0})$ is the Green function of the
graphene lead at the contact site $|\mathbf{0} \rangle$ and
$G_w(1,1)$ is the Green function of the semi-infinite wire at the
site $n=1$.

From Eqs.~(\ref{eq:T}) and (\ref{eq:V}) it is clear  that all matrix
elements of the $T$-matrix outside the $2\times 2$ subspace vanish.
Therefore the Green function, Eq.~(\ref{eq:GF_def}), within the
chain can be expressed in terms of the $T$-matrix of the contact as
\begin{equation}\label{eq:G_t}
G(n,n')=G_w(n,n') + G_w(n,1)T(1,1)G_w(1,n').
\end{equation}
Here $T(1,1) =\langle 1 | \hat T | 1\rangle$ is the $(1,1)$ matrix element of the
$T$-matrix and
\[G_w(n,n')=\langle n | (\epsilon_+ - \hat H_w)^{-1} | n'\rangle \]
is the Green function of an isolated semi-infinite wire.

We use the tight binding model to describe the electron Hamiltonian of the chain,
\begin{equation}\label{eq:H_wire}
    \hat{H}_w= \gamma_w\sum_{n=1}^\infty \Big( u_w |n\rangle \langle n | +
     |n\rangle \langle n+1 |  +
    |n +1\rangle \langle n |  \Big),
\end{equation}
where $\gamma_w$ is the nearest neighbor hopping matrix
element in the wire, and the on-site energy $u_w \gamma_w$
describes the difference in the work functions between graphene and
the carbon chain (in our notations the Fermi energy of the undoped
graphene sheet is set to zero).

The Green function of the semi-infinite wire can be easily
determined, see Appendix \ref{sec:GF_wire},
\begin{equation}\label{eq:G_w}
    G_w(n,n')=\frac{1}{2i\gamma_w\sin{k}}\left(e^{-i
     k|n'-n|}-e^{-ik(n+n')}\right).
    \end{equation}
Here  $k$ is the magnitude of the electron quasimomentum, which is related to the
energy of the electrons by $\epsilon =u_w \gamma_w+2 \gamma_w \cos k$.

With the aid of Eqs.~(\ref{eq:T}), (\ref{eq:V}) and (\ref{eq:G_0}) we can readily
express $T(1,1)$ in Eq.~(\ref{eq:G_t}) in terms of the Green function of the
graphene lead, $G_g(\mathbf{0},\mathbf{0})$. This yields for the combined Green
function evaluated within the wire,
\[
G(n,n')=\frac{e^{-ik(n'-n)}}{2i\gamma_w
 \sin
 k}\left[1-\frac{1-\gamma\, e^{ik}\,
 G_g(\mathbf{0},\mathbf{0})}{1-\gamma\, e^{-ik}\,
 G_g(\mathbf{0},\mathbf{0})} \, e^{-2ikn}\right].
\]
Here we introduced a combination of hopping integrals in the junction and in the
chain, $\gamma=\gamma_c^2/\gamma_w $.

Comparing the last expression with Eq.~(\ref{eq:r_G_wire}) we obtain the
reflection amplitude of the junction,
\begin{equation}\label{eq:reflection_G}
    r(\epsilon)=-\frac{1-\gamma \, e^{ik}\,
    G_g(\mathbf{0},\mathbf{0})}{1-\gamma \, e^{-ik}
    \, G_g(\mathbf{0},\mathbf{0})}.
\end{equation}

Equation (\ref{eq:reflection_G}) is the main result of this
subsection. It expresses the reflection amplitude at the contact in
terms of the Green function of the lead at the contact point with
the wire, $G_g(\mathbf{0},\mathbf{0})$, and holds for an arbitrary
lead.

\subsection{Graphene leads with zigzag edges} \label{sec:lead}

We now specialize to the case, in which the graphene lead is
terminated at the zigzag edge. The zigzag edge has the smallest
number of broken bonds per unit length. It is therefore likely that
the gap which appears in the graphene strip in the experiments of
Ref.~\onlinecite{Marc} is formed along this edge.

We use the nearest neighbor tight binding model to describe the
electron dynamics in graphene and denote the electron $\pi$-orbitals
localized at the atomic sites by $|A, \mathbf{N} \rangle $ (A
sublattice) and $|B, \mathbf{N} \rangle $ (B sublattice). Here
$\mathbf{N}=(N_1,N_2)$ labels the unit cell with a Bravais lattice
vector $N_1 \mathbf{a}_1 +N_2 \mathbf{a}_2$, see Fig.~\ref{fig:1}.
The site $| A, \mathbf{N}=\mathbf{0} \rangle$ is chosen at the atom
which is connected to the carbon chain.

In these notations $G_g(\mathbf{0},\mathbf{0})$  in
Eq.~(\ref{eq:reflection_G}) can be expressed in terms of the Green
function of the semi-infinite graphene plane, $\hat{G}_g
=(\epsilon_+ - \hat H_g)$, as follows:
\begin{equation}\label{eq:G_0_graphene}
    G_g(\mathbf{0},\mathbf{0})=\langle A, \mathbf{N}=\mathbf{0} | \hat{G}_g | A,
    \mathbf{N}=\mathbf{0} \rangle.
\end{equation}

\begin{figure}[ptb]
\includegraphics[width=8.0cm]{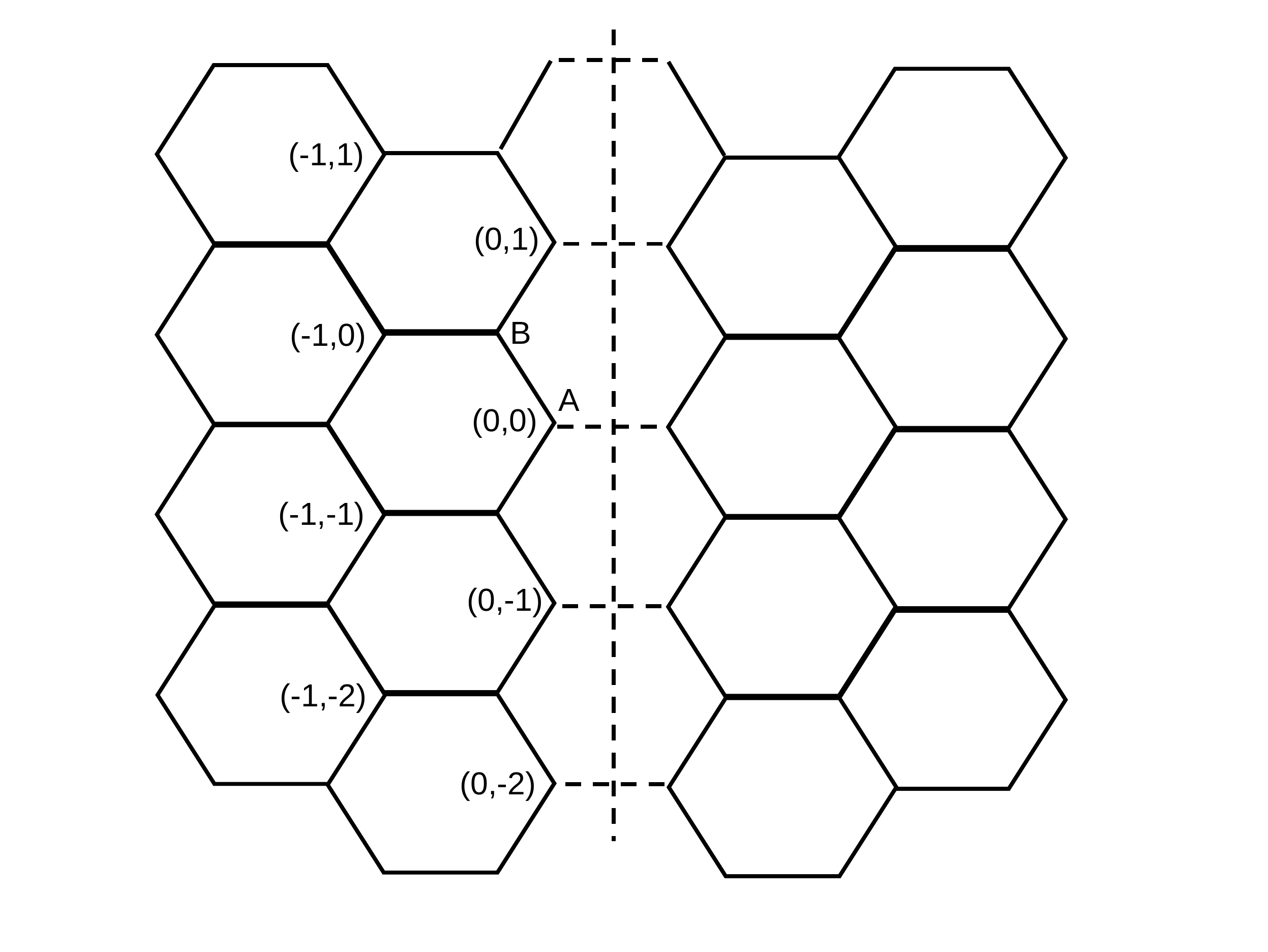}
\caption{An infinite graphene plane is separated into two halves by
adding the perturbation $\hat V_g$, which nullifies the tunneling
along the dashed bonds.} \label{fig:2a}
\end{figure}

In order to evaluate the Green function of the semi-infinite plane
$\hat{G}_g$ we start with the infinite graphene plane and add the
perturbation $\hat V_g$, which nullifies the tunneling through the
bonds which separate the plane into two halves along the zigzag
edge, see Fig.~\ref{fig:2a}.

The Green function of the infinite plane is diagonal in the
quasimomentum representation due to the translation symmetry. We
introduce the spinor Bloch functions as
$\Psi_{\mathbf{k}}^T(\mathbf{N})=\exp(iK_1 N_1+iK_2
N_2)(\psi_A(\mathbf{k}),\psi_B(\mathbf{k}))$, where $\textbf{k}$ is
the quasimomentum,  $K_1=\mathbf{k}\cdot \mathbf{a}_1$ and
$K_2=\mathbf{k}\cdot \mathbf{a}_2$ are the projections of the
quasimomentum onto $\mathbf{a}_1$ and $\mathbf{a}_2$, and
$\psi_{A/B}$ are the wave function amplitudes on the $A/B$
sublattices. In the quasimomentum representation the (inverse) Green
function of an infinite graphene plane can be written as matrix in
the $A/B$ sublattice space,
\begin{equation}\label{eq:G_0g}
G^{-1}_{0,g}(\mathbf{k})=-\gamma_g \left(
  \begin{array}{cc}
    -\varepsilon_+  &\! 1\!+\!e^{i K_{1}}\!+\!e^{-i K_{2} }\!\\
    \!1\!+\!e^{-i K_{1}}\!+\!e^{i K_{2}}\! & -\varepsilon_+ \\
  \end{array}
\right).
\end{equation}
Here $\gamma_g$ denotes the nearest neighbor hopping integral in
graphene and we introduced the dimensionless energy
$\varepsilon_+\equiv \epsilon_+/\gamma_g$.

The Dirac points at the corners of the hexagonal Brillouin zone
correspond to $K_1=K_2=\pm 2\pi /3$. At these points the
off-diagonal matrix elements in the above equation vanish. The
Hamiltonian near these points reduces to the familiar Dirac equation
with the linear spectrum near the Dirac points as shown in
Fig.~\ref{fig:2}.

The perturbation $\hat{V}_g$ that cuts the graphene plane into two
halves is given by (see Fig.~\ref{fig:2a}),
\begin{eqnarray}\label{eq:V_g_cooord}
    \hat{V}_g&=& \gamma_g \delta_{N_2,N_2'}
    \left[u_g \, \delta_{N_1,N_1'} \left(%
\begin{array}{cc}
  \delta_{N_1,0} & 0 \\
  0 &  \delta_{N_1,1}\\
\end{array}%
\right) \right. \nonumber \\
&& - \left. \left(
\begin{array}{cc}
   0 &  \delta_{N_1,0} \delta_{N_1',1}   \\
   \delta_{N_1,1} \delta_{N_1',0} &  0\\
\end{array}%
\right)\right].
\end{eqnarray}
The second matrix in the brackets nullifies electron tunneling between the two
halves of the plane, and the first matrix describes the on-site potential for
the atoms along the zigzag edge. This potential is parameterized in our model by
the dimensionless parameter $u_g$, which is equal to the ratio of the on-site
potential to the hopping integral $\gamma_g$. Because of the diminished number of
neighbors for the edge atoms the on-site potential is expected to be positive and
have a magnitude of the order of $eV$, i.e. of the same order as the hopping
integral, $0< u_g \lesssim 1$.

Due to the symmetry of the problem with respect to translations
along the edge, $(N_1,N_2)\to(N_1, N_2+m)$ the corresponding
quasimomentum, $K_2$, is conserved. Therefore below we use a mixed
position/quasimomentum representation, $(N_1,K_2)$.

In this  representation the matrix $\hat{V}_g $ is independent of $K_2$ and has
nonzero matrix elements only in the $2\times 2$ space spanned by the states
$|A,N_1=0\rangle$ and $|B,N_1=1\rangle$, which correspond to the carbon atoms on
the opposite sides of the divide separating the plane into two halves. In this
$2\times 2$ subspace $\hat{V}_g $ is given by
\begin{equation}\label{eq:V_g_mixed}
    V_g=\gamma_g \left(%
\begin{array}{cc}
  u_g & -1  \\
  -1  & u_g \\
\end{array}%
\right).
\end{equation}

In this  representation the matrix $\hat{V}_g $ is independent of $K_2$ and has
nonzero matrix elements only in the $2\times 2$ space spanned by the states
$|A,N_1=0\rangle$ and $|B,N_1=1\rangle$, which correspond to the carbon atoms on
the opposite sides of the divide separating the plane into two halves. In this
$2\times 2$ subspace $\hat{V}_g $ is given by
\begin{equation}\label{eq:V_g_mixed}
    V_g=\gamma_g \left(%
\begin{array}{cc}
  u_g & -1  \\
  -1  & u_g \\
\end{array}%
\right).
\end{equation}

The reflection amplitude of the junction, Eq.~(\ref{eq:reflection_G}), depends
only on the Green function of the semi-infinite graphene inside the same $2\times
2$ subspace. In the mixed representation the latter satisfies the equation
\begin{equation}\label{eq:Dyson_half_plane}
G_g(K_2)=G_{0,g}(K_2)+G_{0,g}(K_2)V_g G_g(K_2),
\end{equation}
where the perturbation $V_g$ is given by Eq.~(\ref{eq:V_g_mixed}) and
$G_{0,g}(K_2)$ is the unperturbed Green function inside the $2\times 2$ subspace
(in the mixed representation). The latter is evaluated in Appendix
\ref{sec:GF_half_plane} and is given by

\begin{equation}\label{eq:G_0g2}
    G_{0,g}(K_2)=\frac{1}{\gamma_g\sqrt{ab}}\left(%
\begin{array}{cc}
  \varepsilon &  \!1\! -\!C \!+\!\varepsilon^2\! +\!\sqrt{ab}\! \\
  \!1\! -\!C\! +\!\varepsilon^2\! +\!\sqrt{ab}\! & \varepsilon \\
\end{array}%
\right),
\end{equation}
where we introduced the notations

\begin{equation}\label{eq:abC}
C=4 \cos^2\frac{K_2}{2}, \, \, a=(1+\varepsilon)^2 -C, \, \, b=(1-\varepsilon)^2
-C.
\end{equation}
The branch of $\sqrt{ab}$ in Eq.~(\ref{eq:G_0g2}) is determined by analytic
continuation of $\varepsilon$ from the positive imaginary axis, where $\sqrt{ab}$
takes positive real values.

Using Eqs.~(\ref{eq:V_g_mixed}), (\ref{eq:Dyson_half_plane}) and (\ref{eq:G_0g2})
we obtain
\begin{equation}\label{eq:G_g2}
    G_g(K_2)= \frac{2\sqrt{ab}+
    (1-u_g)a +(1+u_g)b}{\gamma_g[(1-u_g)^2 a
-(1+u_g)^2b]} \left(%
\begin{array}{cc}
  1 & 0 \\
  0 & 1 \\
\end{array}%
\right).
\end{equation}
The off-diagonal matrix elements in the above expression vanish, as they should
due to the absence of tunneling between the two half-planes. The $(1,1)$ matrix
element determines the Green function at the zigzag edge for a given
quasimomentum $K_2$ along the edge. Its imaginary part gives the tunneling
density of states into the edge for a given quasimomentum. It arises from two
distinct contributions of the edge and bulk states, which we discuss next.

\subsubsection{Tunneling density of states into the zigzag edge}
\label{sec:tunneling_DOS}

The tunneling density of states at the zigzag edge of graphene is
described by the imaginary part of the diagonal matrix elements in
the Green function Eq.~(\ref{eq:G_g2}). Physically, the density of
states at the edge contains the contributions from the bulk and edge
states. The contribution of the bulk states is described by the
imaginary part of $\sqrt{ab}$ whereas the contribution of the edge
states corresponds to the pole at $(1-u_g)^2 a -(1+u_g)^2b=0$. This
condition defines the spectrum of the edge states,
\begin{equation}\label{eq:edge_spectrum}
\varepsilon=\frac{1+u^2_g-\sqrt{(1+u^2_g)^2+4u^2_g(2\cos{K_2}+1)}}{2u_g}
\end{equation}
with $\cos{K_2}< -\frac{1}{2}$.

\begin{figure}[ptb]
\includegraphics[width=7.0cm]{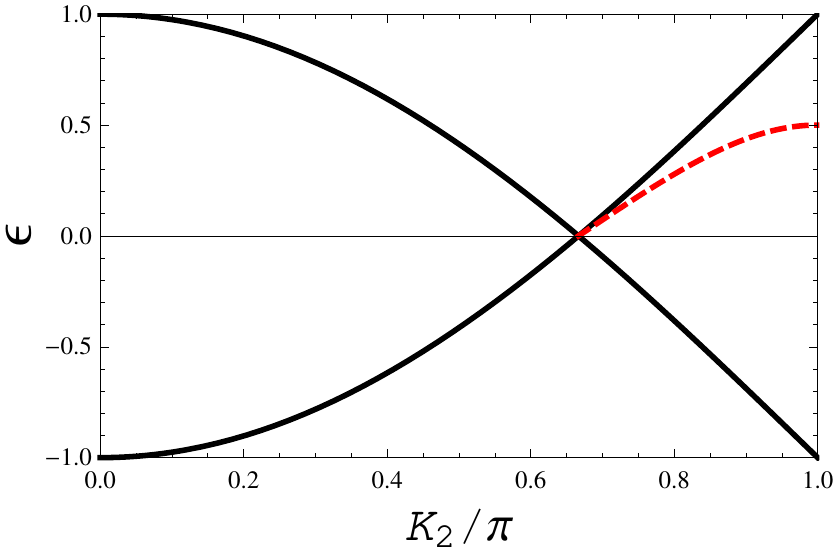}
\caption{Color online: The solid (black) curve represents the
intersection of the bulk state spectrum of graphene with the
$K_1=2\pi/3$ plane that goes through the Dirac point.  The dashed
(red) curve represents the spectrum of the edge states, which exist
only for $2\pi/3<K_2<\pi$. The edge state spectrum lies below the
bulk state spectrum.} \label{fig:2}
\end{figure}

This spectrum is plotted in Fig.~\ref{fig:2}. The inequality
$\cos{K_2}< -\frac{1}{2}$ reflects the fact that for $u_g>0$, the
edge states exist only for $\varepsilon>0$ as shown in the following
text. For $\varepsilon<0$, the density of states for the edge states
vanish, or the numerator in Eq.~(\ref{eq:G_g2}) vanishes together
with the denominator, eliminating the pole.

For weak on-site potential at the edge, $u_g\ll 1$, the edge state spectrum
reduces to $\varepsilon=-u_g(1+2\cos{K_2})$, with $\cos{K_2}\leq -\frac{1}{2}$.
In this limit the spectrum and the wave functions of the edge states can be
understood quite easily. In the absence of the on-site potential at the edge,
$u_g=0$, these states have wave functions which reside only on the $A$-sublattice
and are eigenfunctions of the quasimomentum $\mathbf{k}=(K_1, K_2)$. It is easy
to see from Eq.~(\ref{eq:edge_spectrum}) that these states form a degenerate band
of zero energy states, in agreement with Ref.~\onlinecite{edge_1996}. From
Eq.~(\ref{eq:G_0g}) it follows that in order to obey the Schr\"{o}dinger equation
in the interior of the lead the quasimomentum of such states must satisfy the
condition $e^{-i K_{1}}=- 1-e^{i K_{2}}$. Further, since the wave function of
these states vanishes on the $B$-sublattice they remain eigenstates of the
Hamiltonian even after the plane is separated into two halves. The
normalizability condition for the edge states is $\mathrm{Im} K_1 <0$, implying
$|1+e^{iK_2}|<1$, which is equivalent to the inequality below
Eq.~(\ref{eq:edge_spectrum}). And the amplitude of the edge states decay with a
factor of $2\cos{(K_2/2)}$. For weak on-site potential at the edge, $u_g\ll 1$
the edge state spectrum may be obtained from the first order in perturbation,
$\varepsilon(K_2)=u_g|\psi_A(K_2)|^2$, where $\psi_A(K_2)$ is the wave function
of the edge state at the edge atoms. The normalization condition gives
$|\psi_A(K_2)|^2=1/\sum_{N_1=-\infty}^0 \exp(2N_1 \mathrm{Im} K_1) =-(1+2\cos
K_2)$.

The above consideration illustrates that in the presence of the
on-site potential at the edge the band of edge states acquires a
finite width of order of the on-site potential. For strong on-site
potential at the edge perturbation theory is no longer applicable
and the spectrum of the edge states is given by
Eq.~(\ref{eq:edge_spectrum}). At zero energy the spectrum of these
states is linear, which results in the finite density of states. It
might seem therefore that at small energies, $\varepsilon \ll 1$,
the contribution of the edge states to the tunneling density of
states will be much larger than that of the bulk states. This is not
so however because at small energies near $K_2=2\pi/3$ the wave functions of edge
states extend into the bulk over many lattice spacings, so that the
local density of such states at the edge vanishes linearly with
energy. As a result, for $u_g\sim 1$ the contribution of these
states to the tunneling DoS at the edge turns out to be of the same
order as that of bulk states.

The real space Green function $G_g(\textbf{0},\textbf{0})$ at the
contact point is obtained by integrating the diagonal element of
$G_g(K_2)(\textbf{0},\textbf{0})$ in Eq.~({\ref{eq:G_g2}}) over
$K_2$: $G_g(\textbf{0},\textbf{0})=\int \frac{dK_2}{2\pi}\,
G_g(K_2)$. We write this integral as a contour integral over the
unit circle of the variable $z_2=\exp(iK_2)$. Inside the contour the
integrand has a simple pole corresponding to the edge states and a
branch corresponding to the bulk states. We denote the contribution
of the pole and the branch cut by $G_{pole}$ and $G_{bc}$
respectively,
\begin{eqnarray}\label{eq:G_tilde_result}
  G_g(\mathbf{0},\mathbf{0})=\frac{1}{\gamma_g}(G_{pole}+G_{bc}).
\end{eqnarray}

A lengthy but straightforward calculation gives,
\begin{equation}\label{eq:G_pole}
G_{pole}=-i\frac{|\varepsilon|}{u_g}
    \frac{(1-u^2_g)\theta(\varepsilon)}{\sqrt{3u_g^2-2u_g
    \varepsilon-\varepsilon^2}},
\end{equation}
where $\theta(\epsilon)$ is the step function indicating that the
density of states due to edge states is present only for
$\varepsilon>0$. The contribution of the branch cut, $G_{bc}$, can
be evaluated analytically at low energies,
\begin{widetext}
\begin{eqnarray}\label{eq:G_bc}
G_{bc}&=&-\frac{1}{3u_g}+\frac{\varepsilon}{\pi
u_g}\frac{(1+u^2_g)}{\sqrt{3u_g^2-2u_g
\varepsilon-\varepsilon^2}}\log{\frac{2|\varepsilon|}{\sqrt{3(3u_g^2-2u_g
\varepsilon-\varepsilon^2)}+3u_g-\varepsilon}}
  -ic\frac{|\varepsilon|}{u_g},
\end{eqnarray}
\end{widetext}
where $c =
\frac{1}{2\pi}(1+\frac{\pi}{2}-\frac{2\pi}{3\sqrt{3}})\approx0.22$.
The imaginary parts of both contributions (and with them the
tunneling DoS at the edge) vanish linearly with energy at small
energy.

\subsubsection{Reflection coefficient of the junction at low energies}
\label{sec:reflection_junction}

Substituting the previous
Eqs.~(\ref{eq:G_tilde_result})-(\ref{eq:G_bc}) into
Eq.~(\ref{eq:reflection_G}) we obtain a simple expression for the
reflection coefficient of the junction at low energies,
$|\varepsilon | \ll 1$,
\begin{equation}\label{eq:reflection_low}
    |r(\epsilon)|^2 = 1-\eta \varepsilon
\end{equation}
with
$\eta=-2\sqrt{4-u^2_w}[(1-u^2_g)/\sqrt{3u_g}+c]/(u_g+\frac{1}{9u_g}-\frac{u_w}{3})$.
In this regime the electron wave incident from the carbon chain into
the junction is almost perfectly reflected.

\section{Device conductance at low energies: asymmetric resonances}
\label{sec:conductance_wire}

The strong reflection at the junction at low energies indicates that the
transmission coefficient of the whole device in Eq.~(\ref{eq:transmission_r})
also tends to vanish at small energies except in the vicinity of resonances,
$\cos(2\mathcal{N}k +2\delta_0)=1$. Substituting
$|r_{r/l}(\epsilon)|=|r(\epsilon)|$  from Eq.~(\ref{eq:reflection_low}) into
Eq.~(\ref{eq:transmission_r}) we obtain a simple expression for the transmission
coefficient of the device at low energies,
\begin{equation}\label{eq:transmission_small_epsilon}
    \mathcal{T}(\epsilon)=\left[ 1+
    \frac{\kappa}{\epsilon ^2} [1-\cos(2\mathcal{N}k +2\delta_0)]
    \right]^{-1}.
\end{equation}
Here $\delta_0=2 \arctan(\gamma/3u)$ is the contact scattering phase
shift at zero energy and $\kappa=2/\eta^2$, where $\eta$ is a number
of order unity defined below Eq.~(\ref{eq:reflection_low}).

Expanding the cosine near a resonance energy $\epsilon=\epsilon_0$,
we obtain a simple expression for the transmission coefficient at
small energies,
\begin{equation}\label{eq:transmission_resonance}
    \mathcal{T}(\epsilon)=\left[ 1+
    \alpha \left(1-\frac{\epsilon_0}{\epsilon} \right)^2
    \right]^{-1},
\end{equation}
where $\alpha =\mathcal{N}^2 \kappa/2 $ is a dimensionless parameter. This
reproduces the result (\ref{eq:transmission_resonance_asymmetric}) expected from
qualitative considerations in the case of symmetric coupling. The resonance width
is $\Gamma\sim \epsilon_0/\mathcal{N}$. The resonance shape is strongly
asymmetric and markedly different from that of the Breit-Wigner resonance, as
shown in Fig.~\ref{fig:3}.

\section{Summary and discussion}
\label{sec:discussion}

We studied electron transport through a single atom carbon chain
connected to graphene leads. The simplicity of the hybridization
pattern of electron orbitals in graphene and carbon chains enabled
us to construct an analytically solvable model and thereby gain
physical insight into the essential features of electron conduction
in the system.

Transmission through the device is dominated by scattering at the
contacts between the chain and the lead. For typical temperatures
and doping levels in graphene the current-carrying electron states
have energies much smaller than the band width. At these energies
the contact between the chain and the lead becomes almost perfectly
reflecting. Its reflection amplitude can be expressed in terms of
the Green's function, $G_g(\mathbf{0},\mathbf{0} )$, of the lead at
the atomic site connected to the carbon chain, see
Eq.~(\ref{eq:reflection_G}). In this equation the parameter $\gamma$
describes the strength of coupling between the chain and the lead.
At low electron energies the phase factor $e^{ik}$ may be assumed
energy independent, as it changes appreciably only at energy scales
of order of the band width in the wire. In this regime the energy
dependence of the transmission coefficient is dominated by that of
the density of states in the lead. For graphene leads it becomes
linear, see Eqs.~(\ref{eq:reflection_low}) and
(\ref{eq:contact_transmission_estimate}).

For leads with zigzag edges both the bulk and edge states contribute
to the DoS at the contact point. Due to the difference in the
on-site energy between the atoms at the edge and in the interior of
graphene the band of edge states acquires a finite dispersion. The
spectrum of this band is given by Eq.~(\ref{eq:edge_spectrum}) and
is plotted in Fig.~\ref{fig:2}. Although the edge state spectrum is
linear at small energies its contribution to the local DoS at the
edge is not constant, but rather is linear in the electron energy,
$\sim \epsilon \theta(\epsilon)$. This occurs because the edge state
wave functions extend into the bulk to distances which are inversely
proportional $\epsilon$, as explained in
Sec.~\ref{sec:tunneling_DOS}. As the difference in the on-site
potential between the atoms at the edge and in the interior of the
lead is of the same order as the band width the contribution of edge
states to the DoS is of the same order as that of the bulk states.
Therefore a substantial part of the current through the carbon chain
is propagated into the lead by the edge states. The energy
dependence of the reflection coefficient of the junction is
described by Eq.~(\ref{eq:reflection_low}).

The interference between reflection amplitudes of the left and right
junctions gives rise to the transmission coefficient of the device
described by Eq.~(\ref{eq:transmission_small_epsilon}). Due to the
nearly perfect reflection at the contact the energy dependence of
the transmission coefficient of the interconnect has resonant
character. Near the resonance the transmission coefficient is
described by a simple expression,
Eq.~(\ref{eq:transmission_resonance}).

Our main conclusions, namely the linear energy dependence of the
transmission coefficient of the junction between the chain and the
lead and the shape of the resonance in
Eq.~(\ref{eq:transmission_resonance}) do not depend on many of the
simplifying assumptions of our model.

The linear energy dependence of the junction transmission
coefficient holds if the coupling between the chain and the lead is
energy independent. This assumption is valid as long as the electron
energy is smaller than the inverse propagation time across the
contact and holds for more complicated junctions, e.g. a small
peninsular connecting the chain to the lead. In this case
Eqs.~(\ref{eq:reflection_G}) and
(\ref{eq:transmission_small_epsilon})
 will still hold, provided $\gamma$ and
$\eta$ are replaced by the appropriate parameters describing the
coupling strength between the chain and the lead at low energies.
Similarly, Eq.~(\ref{eq:transmission_resonance}) will also hold
provided the resonance energy and the parameter $\alpha$ are chosen
appropriately. The generalization of the resonance shape to the case
of asymmetric contacts is given by
Eq.~(\ref{eq:transmission_resonance_asymmetric}).

The resonant character of transmission will be preserved even in the
presence of the Coulomb interaction in the wire, as long as wire is
short enough so that the one-dimensional correlation effects can be
neglected. Such a wire will act as a molecule with a single resonant
level participating in transport. For longer wires the
one-dimensional correlations need to be taken into account. In this
respect the Umklapp processes and the formation of Friedel
oscillations near the contact points are especially important. The
study of these effects is left for future work.

We are grateful to M. Bockrath, D. Cobden, J. Lau, J. Rehr, and B.
Spivak for useful discussions. This work was supported by the DOE
grants DE-FG02-07ER46452 (W.C. and A.V.A.) and  DE-FC02-00ER41132
(G.F.B.).

\appendix
\section{Green function of a semi-infinite wire}
\label{sec:GF_wire}

We construct the Green function of the semi-infinite wire from that
of the infinite wire by adding a perturbation that nullifies the
hopping between the two halves.

The retarded Green function of the infinite wire in $k$ space is
diagonal and given by
\begin{equation}
G_{0,w}(k)=(\epsilon-\epsilon_k+i0)^{-1}
\end{equation}
with
\begin{equation}
\epsilon_k =u_w \gamma_w+2 \gamma_w \cos k.
\end{equation}
The real space Green function is obtained by integrating over $k$ as
\begin{equation}
G_0(n,n^\prime)=\frac{1}{2\pi}\int dk G_{0,w}(k)e^{ik(n-n^\prime)}
\end{equation}
which gives
\begin{equation}\label{eq:G_{0,w}}
G_{0,w}(n,n^\prime)=\frac{1}{2i \gamma_w\sin k} \exp(-ik|n-n^\prime|)
\end{equation}
where
$k$ is the magnitude of the electron quasimomentum related to the
energy by $\epsilon =u_w \gamma_w+2 \gamma_w \cos k$.

The perturbation
$\hat{V}_w$ which cuts the wire to two halves has non-vanishing matrix elements only in the $2 \times
2$ subspace spanned by the orbitals with $n=0$ and $n=1$, where it
is given by
\begin{equation}\label{eq:V_w}
\left(
  \begin{array}{cc}
    V_w(0,0) & V_w(0,1) \\
    V_w(1,0) & V_w(1,1) \\
  \end{array}
\right)=-\left(
          \begin{array}{cc}
            0 & \gamma_w \\
            \gamma_w& 0 \\
\end{array}
        \right).
 \end{equation}
The $T$ matrix defined in Eq.~(\ref{eq:T}) is also  nonvanishing
only in the $2\times 2$ subspace and can be expressed solely in
terms of the matrix elements of $\hat{G}_{0,w}$ in the $2\times 2$
space,
\begin{equation}\label{eq:G_wm}
    \left(
             \begin{array}{cc}
               G_{0,w}(0,0) & G_{0,w}(0,1) \\
               G_{0,w}(1,0) & G_{0,w}(1,1)\\
             \end{array}
           \right)
    =\frac{1}{2i
\gamma_w\sin k} \left(%
\begin{array}{cc}
  1 & e^{-ik} \\
  e^{-ik} & 1 \\
\end{array}%
\right).
\end{equation}

Using Eqs.~(\ref{eq:G_{0,w}}), (\ref{eq:V_w}), (\ref{eq:G_wm}),
(\ref{eq:G_T}) and (\ref{eq:T}) we obtain the Green function of the
semi-infinite wire, Eq.~(\ref{eq:G_w}).

\section{Derivation of Eq.~(\ref{eq:G_0g2})}
\label{sec:GF_half_plane}

In this appendix we derive the expression for the unperturbed
graphene Green function within the $2\times 2$ subspace spanned by
the rows of atoms on the opposite sides of the dashed links in
Fig.~\ref{fig:2a}.

The unperturbed Green function in the quasimomentum representation
is obtained by inverting the matrix in Eq.~(\ref{eq:G_0g}),
\[
G_{0,g}(\mathbf{k})=-\frac{1}{D}\left(\begin{array}{cc}
                 -\varepsilon & \! 1\!+\!e^{i K_{1}}\!+\!e^{-i K_{2} }\! \\
                 \! 1\!+\!e^{-i K_{1}}\!+\!e^{i K_{2} }\!& -\varepsilon
               \end{array}\right),
\]
where $D=\gamma_g\left[ \varepsilon^2-(\! 1\!+\!e^{-i
K_{1}}\!+\!e^{i K_{2} }\!)(\! 1\!+\!e^{i K_{1}}\!+\!e^{-i K_{2}
}\!)\right]$.

In the mixed representation the Green function $\hat{G}_{0,g}(K_2)$
in the $2\times2$ subspace of states $|A,N_1=0\rangle$ and
$|B,N_1=1\rangle$, can be obtained by the inverse Fourier transform
of $G_{0,g}(\mathbf{k})$ with respect to $K_1$. An elementary
calculation gives
\begin{eqnarray}
&&\langle A,0|\hat{G}_{0,g}(K_2)|A,0\rangle=\langle B,1|\hat{G}_{0,g}(K_2)|B,1\rangle\nonumber\\
&&=\int \frac{dK_1}{2\pi}
\frac{\varepsilon}{D}=\frac{\varepsilon}{\gamma_g \sqrt{ab}},
\end{eqnarray}
and
\begin{eqnarray}
&&\langle A,0|\hat{G}_{0,g}(K_2)|B,1\rangle=\langle B,1|\hat{G}_{0,g}(K_2)|A,0\rangle ^* \nonumber\\
&&=-\int \frac{dK_1}{2\pi} \frac{ 1\!+\!e^{i K_{1}}\!+\!e^{-i K_{2} }}{D} e^{-iK_1}\nonumber\\
&&=\frac{1}{\gamma_g \sqrt{a b}}\left[1-C+\varepsilon^2+\sqrt{a
b}\right],
\end{eqnarray}
where $a$, $b$ and $C$ are defined in Eq.~(\ref{eq:abC}).

Combining the above matrix elements into one $2\times 2$ matrix we
arrive at Eq.~(\ref{eq:G_0g2}).

\end{document}